\documentclass[sigconf,nonacm]{acmart}

\usepackage{amsmath,booktabs,microtype,multirow}
\usepackage[ruled,vlined,linesnumbered]{algorithm2e}
\usepackage{placeins}
\usepackage{xcolor}
\usepackage[capitalise]{cleveref}

\newtheorem{theorem}{Theorem}
\newtheorem{lemma}{Lemma}
\newtheorem{proposition}{Proposition}

\newcommand{\OPT}{\mathrm{OPT}}
\newcommand{\ind}{\mathbb{I}}
\setcopyright{none}
\title{When Greedy Is Sorting: Exact and Certified Ideal Rankings for Novelty Metrics}

\newcommand{\aff}[1]{\texorpdfstring{\textsuperscript{#1}}{}}
\author{Kehan Liu\aff{1}\texorpdfstring{\qquad}{, }Zhaoyun Chen\aff{1}\texorpdfstring{\qquad}{, }Xianquan Yan\aff{1,2,}}
\authornote{Corresponding author: \texttt{yanx@u.nus.edu}}
\affiliation{%
  \institution{Department of \{\aff{1}Physics, \aff{2}Computer Science\}}
  \city{National University of Singapore}
  \country{Singapore}}

\begin{document}

\begin{abstract}
The ideal ranking that normalizes novelty metrics is NP-hard to compute in
general, so widely used \(\alpha\)-nDCG evaluators approximate its denominator
with a greedy rule that rescans every remaining document at each rank.  We
identify a regime in which this ideal ranking is exact and cheap for both
\(\alpha\)-DCG and intent-aware expected reciprocal rank (ERR-IA).  When each
document is positively judged for at most one intent, each metric reduces to
the classical allocation of a fixed budget across chains of decreasing marginal
gains.  With documents preordered within \(G\) intent groups, a metric-specific
virtual-score transformation followed by a multiway merge, which we call
ChainMerge, returns the globally optimal top-\(K\) ranking in
\(O(G+K\log G)\) time.  Starting from \(n\) globally preordered candidates, the
whole computation costs \(O(n+K\log G)\) instead of the \(\Theta(nK)\) of the
greedy rule and yields the exact \(\alpha\)-IDCG denominator.

When intent judgments overlap, exact \(\alpha\)-DCG optimization is NP-hard
under the standard logarithmic discount and fixed \(\alpha=0.5\), even when
every document has exactly three intents and every intent occurs in exactly two
documents.  For arbitrary overlap, we derive a polynomial-size LP over nested
ranking prefixes whose optimum upper-bounds the ideal score.  Paired with any
feasible ranking, this bound gives an instance-specific certificate of the
ranking's global optimality gap.

Exhaustive enumeration confirms both exact algorithms, and exact dynamic
programming finds the LP bound tight on all but eight of 26,700 sampled
overlapping TREC pools.  In the official TREC Web 2009--2012 judgments, 64.8\%
of 22,084 positive documents serve a single intent, yet only 20 of 198 complete
queries do so throughout.  Beyond this exact regime, the LP certifies 189 of the
198 greedy denominators as globally optimal, and the resulting score intervals
preserve the winner in all eight matched comparison groups and 1,069 of 1,074
strict pairwise orderings.  On 65,238 MIND-small news candidates, ChainMerge
returns identical optimal objectives 17--1,011 times faster than the greedy
rule across the tested groupings and cutoffs.  For \(\alpha\)-DCG, these
results trace the hardness of the ideal ranking to intent overlap.  Without
overlap, the ideal ranking of either metric is a sorted merge, and with overlap,
an LP bound can still certify the greedy denominator.

Code is available at
\href{https://github.com/sarinstein-yan/greedy-is-sorting}{\textcolor{ACMDarkBlue}{github.com/sarinstein-yan/greedy-is-sorting}}.
\end{abstract}


\maketitle
{\let\thefootnote\relax\footnotetext{Accepted at the 20th ACM International Conference on Web Search and Data Mining (WSDM~'27), February 15--19, 2027, Hong Kong, China.}}

\section{Introduction}

\begin{figure*}[t]
\centering
\includegraphics{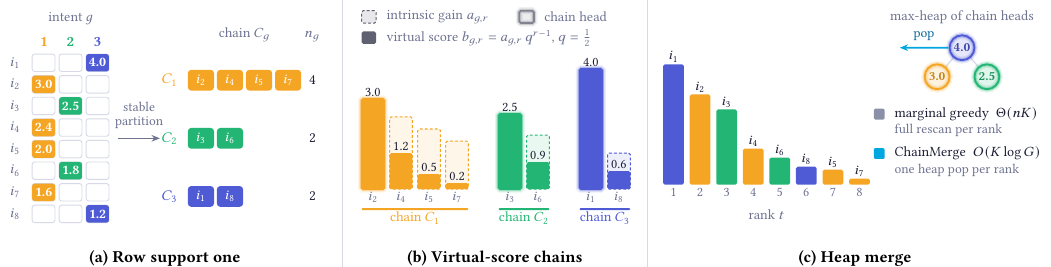}
\Description{Three panels. A document-by-intent matrix in which every document
has one positive intent is partitioned into three chains of four, two, and two
documents. Bar charts show each
chain's intrinsic gains as dashed outlines and its virtual scores as solid bars
that decrease along the chain. A sorted bar chart shows the merged ranking with
a max-heap of the three chain heads.}
\caption{ChainMerge on eight graded documents with three intents and
\(q=\tfrac12\).  (a)~Each document is positive for one intent, so a stable
partition of the globally ordered list yields one chain per intent, with its
\(n_g\) members in gain order.  (b)~The \(r\)-th member keeps the fraction
\(q^{r-1}\) of its intrinsic gain (dashed), so every chain of virtual scores
(solid) is nonincreasing.  Halos mark the chain heads.  (c)~A max-heap over the
chain heads emits the merged ranking, drawn on the scale of (b).  It is optimal under
every nonincreasing discount, and for positive discounts it is also the ranking
that marginal greedy selects.}
\label{fig:overview}
\end{figure*}

Novelty-aware ranking is unusual in that evaluation itself contains a hard
optimization problem.  The numerator of \(\alpha\)-nDCG is computed on a given
list, but its ideal denominator maximizes a history-dependent gain over all
rankings.  Clarke et al. proved this ideal computation NP-hard
\cite{clarke2008novelty}, and Carterette later analyzed the resulting
approximation problem in detail \cite{carterette2011analysis}.  The reference
NIST \texttt{ndeval} implementation therefore builds an ideal ranking by
repeatedly taking the remaining document with the largest marginal gain, which
rescans every unchosen candidate at each rank, and its source labels this
construction a quadratic greedy approximation \cite{ndeval2010}.  We call this procedure \emph{marginal greedy}.
\texttt{pyndeval} inherits the same routine, and \texttt{ir\_measures} inherits
it through \texttt{pyndeval} \cite{pyndeval,macavaney2022irmeasures}.
Intent-aware expected reciprocal rank (ERR-IA) poses a parallel ideal-ranking
problem under a cascade model, because a document's contribution depends on
earlier selections for the same intent.

This worst-case view hides a common structure that admits an exact solution.
Many evaluation instances attach each candidate to one primary group.  For
example, MIND assigns every news article one category and one subcategory
\cite{wu2020mind}.  We define the \emph{row support} of a document as the number
of intents for which it has positive gain, that is, the number of nonzero
entries in its row of the document--intent judgment matrix.  An instance has
\emph{row support one} when every document has row support at most one.  Marginal greedy still rescans every
unchosen document after each selection, even though selecting from one group
changes the gains of no other group.  Within each group, the evolving gains form
one diminishing sequence.

The metric structure turns this sequential computation into a static one.
After the candidates of each intent group are ordered, the \(r\)-th candidate of
the group receives a metric-specific \emph{virtual score}.  For
\(\alpha\)-DCG it is the \(r\)-th largest gain in the group multiplied by
\((1-\alpha)^{r-1}\).  For ERR-IA it is the intent prior times the \(r\)-th
satisfaction probability times the probability that all earlier members of the
group fail to satisfy.  Exchange arguments show that an optimum selects a prefix
of each group and that the virtual scores decrease along that prefix.  What
remains is the classical allocation of a fixed budget across decreasing
marginal returns, so merging the chains is feasible and globally optimal under
every nonincreasing position discount.  We call the resulting algorithm
\emph{ChainMerge}, and \cref{fig:overview} traces it on a small instance.
\emph{Under row support one, ChainMerge returns the exact
\(\alpha\)-IDCG denominator and a globally optimal ERR-IA top-\(K\) ranking,
replacing repeated full scans with an output-sensitive heap merge.}

For \(\alpha\)-DCG, row support one also sits on one side of a sparse
complexity boundary.  Under the evaluator's standard settings, namely fixed
\(\alpha=0.5\), logarithmic discount, and binary judgments, exact optimization
is NP-hard even when every document has exactly three intents and every intent
occurs in exactly two documents.  Tractability therefore holds at row support
one and hardness at row support three, while row support two remains open.  For
arbitrary overlap, we bound the optimum instead of computing it.  Rewriting the
discounted objective as a nonnegative combination of prefix gains (Abel
summation) and relaxing each intent's coverage gain by its concave interpolation
yields a polynomial-size LP, the \emph{prefix LP}, whose value upper-bounds the
exact optimum.  Its gap above any feasible ranking is an instance-specific
optimality certificate.  Keeping one intent per document and bounding the
deleted gains gives a second, solver-free certificate, the \emph{projection
bound}.

This paper makes four contributions.

\begin{itemize}
  \item We prove metric-specific prefix and virtual-score reductions showing
  that the ideal-ranking problems of row-support-one \(\alpha\)-DCG and ERR-IA
  are both instances of classical decreasing-marginal allocation.  ChainMerge is
  therefore globally optimal under arbitrary nonincreasing discounts.
  \item For \(n\) documents, \(G\) intent groups, and cutoff \(K\), ChainMerge
  takes \(O(G+K\log G)\) time from materialized ordered chains and
  \(O(n+K\log G)\) time from a globally preordered list, compared with
  \(\Theta(nK)\) for marginal greedy.  In the binary case, group counts alone
  give the exact \(\alpha\)-IDCG.  An evaluator can check row support while
  reading judgments and switch to ChainMerge whenever it holds.
  \item For overlapping \(\alpha\)-DCG, we prove NP-hardness at row support
  three under fixed standard metric parameters.  For arbitrary overlap, the
  prefix LP and the projection bound provide instance-specific optimality
  certificates.
  \item Exhaustive enumeration and exact dynamic programming confirm the
  reductions and the certificates.  In official TREC judgments, 64.8\% of
  positive documents have row support one, and the LP certifies 189 of 198
  complete-query greedy denominators as globally optimal.  The resulting score
  intervals preserve the winner in all eight matched comparison groups.  On
  MIND-small, ChainMerge is 17--1,011 times faster across the tested groupings
  and cutoffs while preserving the exact objective.
\end{itemize}

\section{Background and Related Work}

\paragraph{Metrics and ideal rankings.}
\(\alpha\)-nDCG was introduced to reward nugget coverage while discounting
redundant gain \cite{clarke2008novelty,clarke2011cascade}.  ERR-IA instead
averages a cascade satisfaction model over intent priors
\cite{chapelle2009err,chapelle2011intent,clarke2011cascade}.  For general
ERR-IA instances, Chapelle et al. proposed both greedy optimization and an exact
branch-and-bound method \cite{chapelle2011intent}.  Exact diversified ranking has
also been formulated through integer programming for an implicit
exemplar-based objective \cite{yu2017concise}.  For the \(\alpha\)-nDCG and
ERR-IA denominators prescribed by the metrics, we identify a polynomial-time
exact regime.  When each document serves at most one intent, ERR-IA admits an
intent-wise prefix reduction and a global chain merge, which replaces the
permutation-tree search of branch and bound with an output-sensitive merge.  The
official TREC Web diversity tracks used NIST's \texttt{ndeval} implementation
\cite{clarke2009web,clarke2010web,clarke2011web,clarke2012web,ndeval2010}, and
modern toolkits expose the same metrics through its backend
\cite{pyndeval,macavaney2022irmeasures}.  Our work concerns the fixed
ideal-ranking problem behind these evaluators.

\paragraph{Diversification and coverage.}
Sequential diversification makes a document's utility depend on the results
already selected.  MMR balances query relevance against redundancy within the
selected set \cite{carbonell1998mmr}, and subtopic retrieval formalized the
resulting departure from independent relevance \cite{zhai2003beyond}.  IA-Select
and xQuAD later made intents explicit and scored marginal intent coverage
\cite{agrawal2009diversifying,santos2010xquad,santos2015survey}.  Subsequent
work developed scalable approximations and specialized objectives for large
result pools \cite{capannini2011efficient}, online methods combining cascade
feedback, position bias, and submodular diversity \cite{hiranandani2020cascade},
and ranking systems learned across queries by directly or differentiably
optimizing diversity objectives
\cite{zhu2015structural,xu2017direct,yu2022optimize,chen2025ma4div,singh2024dial}.
Ideal-ranking computation is the complementary fixed-instance problem, which
optimizes a judged or predicted instance to obtain the metric denominator.  The
same coverage objectives now appear beyond web search.  LANCER reranks RAG
evidence for nugget coverage using \(\alpha\)-nDCG \cite{ju2026lancer}, and
systems based on query decomposition evaluate retrieval under the same metric
\cite{petcu2026query}.

\paragraph{Resource allocation.}
Classical discrete resource allocation assigns a fixed budget across activities
with diminishing marginal returns.  Marginal analysis for single-budget discrete
allocation dates to Fox \cite{fox1966discrete}.  The greedy rule is optimal for
separable concave objectives and extends to polymatroid constraints
\cite{federgruen1986greedy,ibaraki1988resource}.  This theory supplies the merge
primitive in our algorithm.  \emph{The new step is metric specific, namely
showing that \(\alpha\)-DCG and ERR-IA each generate one nonincreasing
virtual-score chain per intent}, with the ranking cutoff as the allocation
budget and the virtual scores as its marginal returns.  In particular, the
ERR-IA reduction proves that an optimum selects a prefix of each intent's
satisfaction order despite the cascade survival products.

\paragraph{Hardness and certification.}
Beyond the general NP-hardness of ideal novelty ranking and its
approximation analysis \cite{clarke2008novelty,carterette2011analysis},
classical greedy approximation for monotone submodular set maximization concerns
an unordered cardinality constraint \cite{nemhauser1978analysis}.  Approximation
algorithms and LP relaxations for ordered objectives include diversified search
ranking \cite{bansal2010approx}, submodular cover-time ranking
\cite{azar2011ranking}, and submodular ranking under prefix budgets
\cite{zhang2022submodular}.  Our hardness result fixes the standard logarithmic
discount, \(\alpha=0.5\), and binary judgments while restricting every document
to three intents and every intent to two documents.  Together with the exact
row-support-one algorithm, this narrows the sparse tractability frontier, and row
support two remains open.  For arbitrary overlap, the prefix LP upper-bounds the
ideal \(\alpha\)-DCG denominator, so its gap above any feasible ranking is an
instance-specific optimality certificate rather than a worst-case approximation
ratio.  The projection bound gives a further certificate without an LP solver.

\section{Problem Setting}

Let \(D\) be a set of \(n\) candidates, \(\mathcal G=\{1,\ldots,G\}\) the
intents, and \(K\le n\) the cutoff.  Discounts satisfy
\(d_1\ge\cdots\ge d_K\ge0\), and standard DCG uses
\(d_t=1/\log_2(t+1)\) \cite{jarvelin2002cumulated}.  A ranking
\(\pi=(\pi_1,\ldots,\pi_K)\) lists \(K\) distinct documents.  An instance has
row support one when each candidate has positive gain for at most one intent.
After dropping zero-gain candidates, row support one partitions the
positive-gain set \(D_+\) into groups \(C_g\), one per intent.  We replace
\((D,K)\) by \((D_+,\min\{K,|D_+|\})\), because omitted zero-gain documents can
fill any remaining ranks without changing the objective.  Below, \(D\),
\(n=|D|\), and \(K\) refer to this reduced instance.

For \(i\in C_g\), let \(a_i\ge0\) be its intrinsic gain, and set
\(q=1-\alpha\in[0,1]\), with the convention \(0^0=1\).  The partitioned
\(\alpha\)-DCG objective is
\begin{equation}
F_\alpha(\pi)=\sum_{t=1}^K d_t a_{\pi_t}
q^{N_{g(\pi_t)}(t-1)},                                      \label{eq:alpha}
\end{equation}
where \(N_g(t-1)=|\{s<t:g(\pi_s)=g\}|\) counts the earlier positions assigned to
group \(g\).  Binary TREC judgments use \(a_i=1\).  In the graded convention used
below, \(a_i\) supplies the intrinsic utility, while the exponent counts
preceding positive judgments for the intent.

For ERR-IA, group \(g\) has prior \(p_g\ge0\), with \(\sum_g p_g=1\), and
candidate \(i\in C_g\) has satisfaction probability \(s_i\in[0,1]\):
\begin{equation}
F_{\mathrm{ERR}}(\pi)=\sum_{t=1}^K d_t p_{g(\pi_t)}s_{\pi_t}
\prod_{h<t:g(\pi_h)=g(\pi_t)}(1-s_{\pi_h}).                 \label{eq:err}
\end{equation}
\emph{Both objectives are history dependent, since the gain of a document
depends on the earlier documents that share its intent.}  \Cref{sec:chain} removes this
dependence under row support one, and \cref{sec:overlap} treats overlapping
intents.

\section{Chain Merging Is Exact under Row Support One}\label{sec:chain}

We first isolate the scheduling fact behind classical decreasing-marginal
allocation, then reduce each metric to it.

\begin{lemma}[Monotone-chain rearrangement]\label{lem:chains}
Suppose feasible items form predecessor chains, each with nonincreasing weights
\(b_{g,1}\ge b_{g,2}\ge\cdots\ge0\).  For nonincreasing nonnegative discounts,
sorting all items by \(b\), with ties respecting chain order, and taking the
first \(K\) maximizes \(\sum_{t\le K}d_tb_{\pi_t}\).
\end{lemma}
\begin{proof}
Without chain constraints, the rearrangement inequality pairs the \(K\) largest
weights with the largest discounts in nonincreasing order.  Because every chain
is itself nonincreasing, this global order never places a successor before its
predecessor.  The unconstrained optimum is therefore feasible.
\end{proof}

\subsection{\(\alpha\)-DCG}

Within each group write \(a_{g,1}\ge\cdots\ge a_{g,n_g}\) and define
\begin{equation}
b^{\alpha}_{g,r}=a_{g,r}q^{r-1}.                             \label{eq:avirtual}
\end{equation}

\begin{theorem}[Exact \(\alpha\)-ChainMerge]\label{thm:alpha}
For every \(\alpha\in[0,1]\), every nonincreasing nonnegative discount, and
every cutoff, globally merging the chains \(b^\alpha_{g,r}\) gives a global
maximizer of \cref{eq:alpha}.
\end{theorem}
\begin{proof}
Fix the positions \(t_1<\cdots<t_m\) occupied by group \(g\).  Their coefficients
\(d_{t_r}q^{r-1}\) are nonnegative and decrease with \(r\).  Rearrangement
therefore transforms any selection without loss into the prefix
\(a_{g,1},\ldots,a_{g,m}\), assigned in decreasing gain order.  Hence a global
maximizer exists with this prefix form.  Once these within-group orders are
fixed, occurrence \((g,r)\) contributes \(d_t b^\alpha_{g,r}\), and
\[
b^\alpha_{g,r+1}=a_{g,r+1}q^r
\le a_{g,r}q^r
\le a_{g,r}q^{r-1}=b^\alpha_{g,r}.
\]
This holds throughout \(q\in[0,1]\), including \(q=0\) under the convention
\(0^0=1\).  Thus each chain decreases and \cref{lem:chains} applies.
\end{proof}

\paragraph{A worked instance.}
\Cref{fig:overview} applies the theorem to eight graded documents, three
intents, and \(q=\tfrac12\).  Each document is positive for one intent, so a
single pass over the globally ordered list splits it into the chains
\(C_1,C_2,C_3\) (\cref{fig:overview}a).  Within a chain, the \(r\)-th member
keeps only the fraction \(q^{r-1}\) of its intrinsic gain, because \(r-1\)
earlier members already cover its intent.  For example, document \(i_4\) has gain
\(2.4\) but virtual score \(1.2\) as the second member of \(C_1\)
(\cref{fig:overview}b).  The chains are nonincreasing, and merging them gives
the ranking in \cref{fig:overview}c, in which the three intents interleave as
\(3,1,2,1,2,3,1,1\).

\paragraph{Marginal greedy is exact under every tie-break.}
At rank \(t\), a remaining member \(i\in C_g\) adds \(d_ta_iq^{N_g(t-1)}\), so
all members of a group share the multiplier \(d_tq^{N_g(t-1)}\).  While the
largest marginal gain is positive,\footnote{Neither the discounts nor the
multipliers increase over the ranks, so once the largest marginal gain is zero,
every later addition is zero and any completion is optimal.} the chosen group's
multiplier is positive, so greedy takes the largest remaining intrinsic gain of
that group, and its addition is \(d_t\) times the group's chain head.  By
induction, these selections form a prefix of a ChainMerge order, with ties
among equal gains ordered consistently, and \cref{lem:chains} makes every
choice among tied heads optimal.  \emph{Marginal greedy is thus exact under row support one for every
tie-breaking rule, and its repeated scans are unnecessary.}  When ties are
broken as in ChainMerge, the two procedures return the same ranking, which is
the sense in which greedy is sorting.

\paragraph{Cost by input representation.}
\Cref{alg:merge} stores only the head of every chain.  If the within-group
arrays are already materialized and ordered by intrinsic gain, the merge uses
\(O(G+K\log G)\) time and \(O(G)\) extra memory.\footnote{We read \(\log G\) as
\(\max\{1,\log G\}\), so the bounds also count the \(K\) heap operations when
\(G=1\).}  When the retrieval list is
sorted by intrinsic gain, a stable partition, which appends each document to its
group's array in list order, materializes the arrays in \(O(n)\) time and gives
an \(O(n+K\log G)\) path from scratch.  Unsorted groups require
\(O(n\log n)\) preprocessing.  With binary judgments, every chain is the
geometric sequence \(1,q,q^2,\ldots\) truncated after \(n_g\) terms, so the
merged scores, and hence the exact \(\alpha\)-IDCG, depend only on the group
counts \(n_g\).  These counts can be accumulated while judgments are read.
\Cref{tab:complexity} summarizes the cost of each input representation.

\begin{table}[t]
\caption{Worst-case time by input representation. Global-list paths include
the stable partition or the counting pass, and materialized paths start from
ordered group arrays or group counts.}
\label{tab:complexity}
\centering
\small
\begin{tabular}{lcc}
\toprule
Method / input & Time & Extra space\\
\midrule
Marginal greedy & \(\Theta(nK)\) & \(O(n+G)\)\\
ChainMerge, unsorted & \(O(n\log n+K\log G)\) & \(O(n+G)\)\\
Global preordered list & \(O(n+K\log G)\) & \(O(n+G)\)\\
Materialized chains & \(O(G+K\log G)\) & \(O(G)\)\\
Binary list, count-only & \(O(n+G+K\log G)\) & \(O(G)\)\\
Binary group counts & \(O(G+K\log G)\) & \(O(G)\)\\
\bottomrule
\end{tabular}
\end{table}

\begin{algorithm}[t]
\caption{ChainMerge for materialized, ordered chains}\label{alg:merge}
\KwIn{chains \(b_{g,1}\ge\cdots\), cutoff \(K\)}
initialize a max-heap with \((b_{g,1},g,1)\) for each nonempty group\;
\For{\(t=1,\ldots,K\)}{
  pop maximum \((b_{g,r},g,r)\), and output candidate \((g,r)\)\;
  \If{\(r+1\le n_g\)}{push \((b_{g,r+1},g,r+1)\)\;}
}
\end{algorithm}

\subsection{ERR-IA}

Within group \(g\), sort \(s_{g,1}\ge\cdots\ge s_{g,n_g}\) and define
\begin{equation}
b^{\mathrm{ERR}}_{g,r}=p_gs_{g,r}\prod_{h<r}(1-s_{g,h}).      \label{eq:evirtual}
\end{equation}
When chain \(g\) is ranked in this order, its virtual score is the
prior-weighted probability that the \(r\)-th member is the first document to
satisfy intent \(g\), because the \(r-1\) members before it must all fail.

\begin{theorem}[Exact ERR-IA ChainMerge]\label{thm:err}
Merging the chains in \cref{eq:evirtual} globally maximizes \cref{eq:err} for
every nonincreasing nonnegative discount.
\end{theorem}
\begin{proof}
First fix the selected candidates of intent \(g\).  Consider two consecutive
occurrences of this intent in the ranking, at global positions \(t<t'\), with
satisfaction probabilities \(x\) and \(y\).  Let
\(P=p_g\prod_{h<t:g(\pi_h)=g}(1-s_{\pi_h})\) be their prior-weighted prefix
survival.  No occurrence of \(g\) lies between them.  Placing \(x\) before \(y\)
rather than \(y\) before \(x\) changes their combined contribution by
\(P(x-y)(d_t-d_{t'})\).  Contributions of other intents are unchanged, and
survival after both is the same.  Eliminating adjacent inversions therefore
orders the selected probabilities nonincreasingly within every intent.

It remains to establish prefix selection.  Suppose an occurrence at global
position \(t\) has satisfaction \(s\) and prior-weighted prefix survival \(P\).
Let \(V\) be the conditional expected discounted contribution from the future
first satisfaction of this intent, given that the current occurrence does not
satisfy.  The contribution from the current occurrence and this suffix is
\(P[sd_t+(1-s)V]\).  Future first-satisfaction events have total probability at
most one and occur at discounts no larger than \(d_t\), so \(0\le V\le d_t\).
Replacing \(s\) by an unselected \(s'>s\) changes the objective by
\[
P(s'-s)(d_t-V)\ge0.
\]
Repeated replacement, followed by the within-intent ordering above, transforms
an optimum without loss into one that selects a prefix of each satisfaction
order.

Finally, writing \(Q_{g,r}=p_g\prod_{h<r}(1-s_{g,h})\ge0\),
\[
b^{\mathrm{ERR}}_{g,r+1}
=Q_{g,r}s_{g,r+1}(1-s_{g,r})
\le Q_{g,r}s_{g,r}=b^{\mathrm{ERR}}_{g,r}.
\]
Thus every virtual-score chain is nonincreasing, and \cref{lem:chains} applies.
\end{proof}

The greedy argument given after \cref{thm:alpha} carries over with the group
multiplier \(d_tp_g\prod_{h<t:g(\pi_h)=g}(1-s_{\pi_h})\) in place of
\(d_tq^{N_g(t-1)}\), so marginal greedy is exact for ERR-IA under every
tie-break as well.  \emph{A count-based metric and a probabilistic cascade metric
thus share the same chain structure, despite their different functional
forms.}

\section{Hardness and Certificates under Intent Overlap}\label{sec:overlap}

\begin{figure*}[t]
\centering
\includegraphics{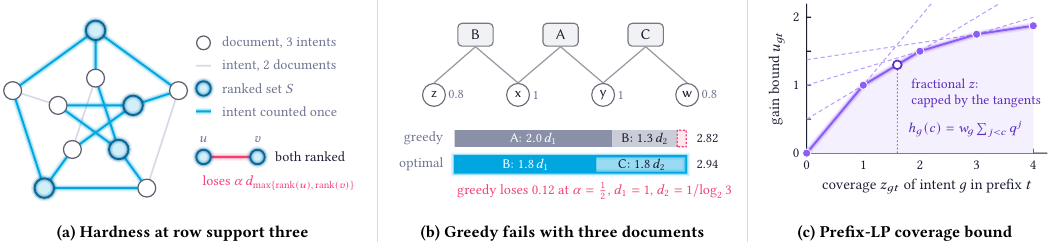}
\Description{Three panels. The Petersen graph with four highlighted vertices
forming an independent set and their twelve incident edges highlighted. A
bipartite diagram of three documents and four intents with two horizontal bars
comparing the greedy and optimal rankings. A concave piecewise-linear curve with
dashed tangent lines and a shaded region below it.}
\caption{Intent overlap breaks the chain structure.  (a)~In the reduction from
Independent Set on cubic graphs, documents are vertices and intents are edges.
A ranked independent set counts every touched intent once and reaches the cap
\(3\sum_td_t\), while ranking both endpoints of an edge loses \(\alpha\) times
the later discount.  (b)~With \(K=2\) and \(\alpha=\tfrac12\), greedy ranks the
heaviest document \(\mathsf{A}\) first and ends 0.12 below the optimal ranking
\((\mathsf{B},\mathsf{C})\).
Bar lengths are discounted contributions, darker for rank~1 and lighter for
rank~2.  (c)~For one intent with \(w_g=1\), \(q=\tfrac12\), and \(m_g=4\)
covering documents, the prefix LP caps the coverage gain \(u_{gt}\) by every
tangent (dashed) of the concave interpolation of \(h_g\).  The cap equals
\(h_g(c)\) at integral coverage (filled) and interpolates at fractional coverage
(hollow).}
\label{fig:overlap}
\end{figure*}

For an overlapping binary incidence matrix \(A\in\{0,1\}^{n\times G}\) and
intent weights \(w_g\ge0\), let
\(N_g^\pi(t-1)=\sum_{s<t}A_{\pi_sg}\).  The objective in this section is
\begin{equation}
F_\alpha(\pi)=\sum_{t=1}^K d_t\sum_g
A_{\pi_tg}w_gq^{N_g^\pi(t-1)}.                                \label{eq:overlap}
\end{equation}
It reduces to \cref{eq:alpha} under row support one, and the hardness result
uses unit intent weights.

\subsection{Hardness at standard parameters}

When a document covers several intents, selecting it advances several chains at
once, so neither static virtual scores nor \cref{lem:chains} applies.

\begin{theorem}[Sparse hardness at fixed standard parameters]\label{thm:hard}
For every fixed \(\alpha\in(0,1]\), exact binary \(\alpha\)-DCG maximization with
standard logarithmic discount is NP-hard, even when every document has exactly
three intents and every intent occurs in exactly two documents.
\end{theorem}
\begin{proof}
Reduce from Independent Set on cubic graphs, which is NP-hard
\cite{alimonti2000cubic}.  A vertex becomes a document and an edge becomes an
intent, covered by its two endpoint documents.  For a ranked vertex set \(S\) of
size \(K\), every vertex initially contributes three units.  If both endpoints
of edge \(e\) are selected, its later endpoint contributes \(1-\alpha\) instead
of one.  Therefore the exact identity is
\begin{equation}
F_\alpha(\pi)=3\sum_{t=1}^K d_t-
\alpha\!\sum_{e=\{u,v\}\in E[S]}
d_{\max\{\mathrm{rank}_\pi(u),\mathrm{rank}_\pi(v)\}}.       \label{eq:hardidentity}
\end{equation}
All discounts and \(\alpha\) are positive, so the explicit cap
\(3\sum_td_t\) is reached if and only if \(S\) is independent.  An exact ranking
oracle therefore returns an independent \(K\)-set exactly when one exists.
Checking the returned set is combinatorial, so the reduction requires no
arithmetic threshold comparison between logarithmic discounts.  The
construction has the stated 3-by-2 incidence.
\end{proof}

\Cref{fig:overlap}a shows the construction on the Petersen graph, which is
cubic.  Every intent touched by the ranked independent set is counted once,
while ranking both endpoints of an edge forfeits \(\alpha\) times the discount of
the later endpoint.  The exclusion of \(\alpha=0\) is necessary, because
redundancy then disappears and the objective is modular.  \emph{History
dependence alone therefore does not make ideal novelty ranking hard.  Hardness
enters once documents cover several intents, already at row support three.}

\paragraph{Greedy fails with three documents.}
Overlap also breaks the exactness of marginal greedy, and three documents
suffice.  With two documents and \(K=2\), strict failure is impossible.  Let
\(W\) be the total weight of their shared intents and \(X\) and \(Y\) the
weights exclusive to each document.  Greedy ranks first the document with the
larger exclusive weight,\footnote{If \(d_1=0\), both rankings have value zero.} say \(X\ge Y\), and the two complete rankings differ by
\((d_1-d_2)(X-Y)\ge0\) in its favor, because the shared contributions cancel.
\Cref{fig:overlap}b gives a three-document failure under the evaluator's
defaults \(\alpha=0.5\), \(d_1=1\), and \(d_2=1/\log_23\).  Intents
\(\mathsf{x}\) and \(\mathsf{y}\) have weight 1, intents \(\mathsf{z}\) and
\(\mathsf{w}\) have weight 0.8, and the documents cover
\(\mathsf{A}=\{\mathsf{x},\mathsf{y}\}\), \(\mathsf{B}=\{\mathsf{x},\mathsf{z}\}\),
and \(\mathsf{C}=\{\mathsf{y},\mathsf{w}\}\).  Greedy strictly prefers
\(\mathsf{A}\), whose gain 2 exceeds 1.8.  Every second choice then repeats
\(\mathsf{x}\) or \(\mathsf{y}\) and adds only 1.3, so greedy attains
\(2+1.3\,d_2\approx2.820\).  The ranking \((\mathsf{B},\mathsf{C})\) covers four
intents once each and attains
\(1.8+1.8\,d_2\approx2.936\).  \emph{Overlapping queries therefore need
certificates rather than an exactness guarantee.}

\subsection{Prefix-LP certificate}\label{sec:lp}

Because exact optimization is NP-hard under overlap, we instead bound the
optimum with a polynomial-size LP.  The cumulative novelty gain of a ranking
prefix depends only on how many of its documents cover each intent, with
diminishing returns for repeated coverage.  Abel summation expresses the
discounted score as a nonnegative combination of these prefix gains, and the LP
bounds them jointly.

We first give \cref{eq:overlap} an exact prefix representation.  We retain only
intents covered by at least one document, so that
\(m_g=|\{i:A_{ig}=1\}|\ge1\).  Empty intents contribute nothing to the
objective.  Define the concave geometric coverage function
\begin{equation}
h_g(r)=w_g\sum_{j=0}^{r-1}q^j,
\qquad c_g(S)=\sum_{i\in S}A_{ig}.                            \label{eq:h}
\end{equation}
For the ranking prefix \(S_t=\{\pi_1,\ldots,\pi_t\}\), the cumulative novelty
gain before position discount is exactly \(\Gamma_t=\sum_gh_g(c_g(S_t))\),
because each intent contributes occurrence marginals \(w_g,w_gq,w_gq^2,\ldots\)
regardless of interleaving.  Set \(d_{K+1}=0\) and \(\Delta_t=d_t-d_{t+1}\ge0\).
Abel summation gives
\begin{equation}
F_\alpha(\pi)=\sum_{t=1}^K d_t(\Gamma_t-\Gamma_{t-1})
=\sum_{t=1}^K\Delta_t\Gamma_t.                             \label{eq:abel}
\end{equation}
Because every \(\Delta_t\) is nonnegative, upper bounds on the prefix gains
\(\Gamma_t\) combine into an upper bound on the full objective.  The LP below
bounds all prefix gains jointly while keeping the prefixes nested.

Linearly interpolate \(h_g\) between its integer points.  Its segment slopes are
\(w_gq^r\), so the interpolation is concave and its hypograph is the
intersection of tangent halfspaces, drawn in \cref{fig:overlap}c.
Introduce \(x_{it}\), the fractional membership of document \(i\) in the
length-\(t\) prefix, and \(u_{gt}\), which stands for the coverage gain of intent
\(g\) in that prefix.  Write
\(z_{gt}(x)=\sum_iA_{ig}x_{it}\).  The prefix LP is
\begin{align}
U_{\mathrm{LP}}=\max\quad
 &\sum_{t=1}^K\Delta_t\sum_g u_{gt}                         \label{eq:lpobj}\\
\text{s.t.}\quad
 &\sum_i x_{it}=t && t=1,\ldots,K,                          \label{eq:lpcard}\\
 &0\le x_{i1}\le\cdots\le x_{iK}\le1 && i\in D,            \label{eq:lpnest}\\
 &0\le u_{gt} && g,t,\\
 &u_{gt}\le h_g(r)+w_gq^r\!\left(z_{gt}(x)-r\right)
 && g,t,\ 0\le r<m_g.                                      \label{eq:lphypo}
\end{align}
In this LP, \cref{eq:lpcard} fixes the size of each prefix, and
\cref{eq:lpnest} keeps the prefixes nested, so a document that enters a prefix
stays in every longer one.  \Cref{eq:lphypo} caps the coverage gain of every
intent by each tangent of its interpolation, and at an integral count
\(z_{gt}=c\) the tightest of these caps is exactly \(h_g(c)\).  The LP has
\(O(nK+GK)\) variables and \(O(nK+K\sum_gm_g)\) constraints.\footnote{For
rational \(q\), intent weights, and discounts, the coefficient encoding length is
also polynomial.}

\begin{theorem}[Global \(\alpha\)-DCG overlap certificate]\label{thm:cert}
For arbitrary binary overlap, every \(\alpha\in[0,1]\), and every nonincreasing
nonnegative discount,
\[
\OPT:=\max_\pi F_\alpha(\pi)\le U_{\mathrm{LP}}.
\]
Thus any feasible ranking \(\widehat\pi\) has the deterministic certificate
\[
0\le\OPT-F_\alpha(\widehat\pi)
\le U_{\mathrm{LP}}-F_\alpha(\widehat\pi).
\]
\end{theorem}
\begin{proof}
Embed any integral ranking by setting \(x_{it}=\ind\{i\in S_t\}\).  Cardinality
and nesting hold.  At the integer count \(c_g(S_t)\), set
\(u_{gt}=h_g(c_g(S_t))\).  All tangent inequalities hold with the concave
envelope, and \cref{eq:abel} makes the LP objective exactly the ranking value.
Every ranking is therefore LP-feasible, so the maximum LP value upper-bounds all
of them.  Subtract any feasible ranking value.
\end{proof}

The bound is tight whenever the LP has an optimum with \(x_{it}\in\{0,1\}\).
Such nested prefixes define a ranking \(\pi\),\footnote{The variables
\(u_{gt}\) need not be integral, and their caps need not bind when
\(\Delta_t=0\).} and
\[
U_{\mathrm{LP}}\le\sum_{t=1}^K\Delta_t\sum_g h_g\bigl(z_{gt}(x)\bigr)
=F_\alpha(\pi)\le\OPT.
\]
A strict gap \(U_{\mathrm{LP}}>\OPT\) therefore requires
fractional document memberships in every LP optimum, so that some document is
split across prefix positions and the interpolation credits coverage that no
single ranking attains.

\begin{proposition}[Graded overlap envelope]\label{prop:graded}
Suppose every incidence \((i,g)\) carries gain \(a_{ig}\ge0\), with count-based
decay \(q\).  Sort the gains incident to \(g\) as
\(a_{g,1}^{\downarrow}\ge a_{g,2}^{\downarrow}\ge\cdots\), and define
\[
H_g(0)=0,\qquad
H_g(r)=\sum_{j=1}^{r}a_{g,j}^{\downarrow}q^{j-1}\quad (r\ge1).
\]
Replacing \cref{eq:lphypo} by the tangent constraints
\begin{equation}
u_{gt}\le H_g(r)+a_{g,r+1}^{\downarrow}q^r
\bigl(z_{gt}(x)-r\bigr),
\qquad 0\le r<m_g,                                             \label{eq:gradedhypo}
\end{equation}
gives an upper bound on the optimal graded count-decay objective.
At an integral prefix, the envelope is exact when its selected \(g\)-incidences
are the largest available gains and occur in nonincreasing gain order.
\end{proposition}
\begin{proof}
If a prefix contains \(r\) incidences for intent \(g\), rearrangement bounds its
cumulative contribution by pairing the \(r\) largest incident gains with
\(1,q,\ldots,q^{r-1}\), yielding \(H_g(r)\).  The slopes
\(a_{g,r+1}^{\downarrow}q^r\) are nonincreasing, so the interpolation is concave
and has the same tangent representation as \cref{eq:lphypo}.  Abel summation
combines the prefix bounds with nonnegative coefficients \(\Delta_t\), and the
LP relaxation preserves the upper bound.  The stated prefix condition attains
the rearrangement bound.
\end{proof}

\paragraph{Projection bound without an LP solver.}
Write \(a_{ig}=w_gA_{ig}\) in the binary case.  For every positive document,
choose one retained intent \(\phi(i)\in\{g:a_{ig}>0\}\), keep only that
incidence, and solve the resulting row-support-one instance exactly.  Let
\(\OPT_\phi\) denote its ChainMerge optimum.  For intent \(g\), let
\(R_g=\{a_{ig}:a_{ig}>0,\ \phi(i)\ne g\}\) be the multiset of deleted gains,
sorted as \(r_{g,1}\ge r_{g,2}\ge\cdots\).  Then
\begin{equation}
\begin{aligned}
\OPT &\le\OPT_\phi\\
&\quad+\sum_g\sum_{j=1}^{\min\{|R_g|,K\}}d_jr_{g,j}q^{j-1}.
\end{aligned}                                                \label{eq:residual}
\end{equation}
To see the bound, fix any ranking.  Removing deleted incidences weakly reduces
the preceding count for every retained incidence, so the retained contribution
is no larger than the value of the same ranking in the projection and hence no
larger than \(\OPT_\phi\).  Enumerate the ranked deleted incidences of intent
\(g\) in occurrence order.  Its \(j\)-th such incidence has global position at
least \(j\) and at least \(j-1\) preceding occurrences of \(g\), so its
coefficient is at most \(d_jq^{j-1}\).  The rearrangement inequality pairs these
decreasing bounds with the largest deleted gains, yielding the residual term in
\cref{eq:residual}.  Summing over intents proves the claim.  We call the
right-hand side the \emph{projection bound}.  It requires no solver and holds for
every choice of \(\phi\), so minimizing it over \(\phi\) can only tighten the
certificate.

\section{Empirical Evaluation}

We evaluate the exact and certified regimes through four research questions.
RQ1 asks whether ChainMerge recovers the exact optimum, RQ2 how often row support
one occurs in official data, RQ3 how tight the overlap certificates are, and RQ4
whether measured runtimes reflect the predicted complexity separation.

\subsection{Implementation and Protocol}

\paragraph{Existing evaluators.}
We examined the code of five evaluators, which are NIST \texttt{ndeval}, its
\texttt{pyndeval} and PT-Ranking ports, \texttt{ir\_measures}, and RankSys
\cite{ndeval2010,pyndeval,yu2020ptranking,macavaney2022irmeasures,ranksys}.
NIST's reference routine builds the ideal ranking one position at a time by
rescanning every unplaced document \cite{ndeval2010}.  Its ports preserve this
procedure, and \texttt{ir\_measures} delegates diversity metrics to
\texttt{pyndeval} \cite{macavaney2022irmeasures}.  None of them detects row
support one or merges marginal chains.

\paragraph{Exact recovery.}
In each of three independent replicates, we generate 1,000 row-support-one
\(\alpha\)-DCG instances and 1,000 row-support-one ERR-IA instances.  We draw
\(n\) uniformly from \(\{3,\ldots,8\}\), assign each item independently to one
of at most four intents, and draw \(K\) uniformly from
\(\{1,\ldots,\min(5,n)\}\).  For \(\alpha\)-DCG, gains are uniform on
\([0.01,4]\) and \(\alpha\) is uniform on \([0,1]\).  For ERR-IA, satisfaction
probabilities are uniform on \([0.001,0.999]\), while intent weights are
independent uniforms on \([0.01,1]\) normalized to sum to one.  We compare
ChainMerge with exhaustive enumeration of every ordered top-\(K\) list.

\paragraph{TREC Web diversity.}
We parse all 198 queries in the official 2009--2012 Web diversity qrels
\cite{clarke2009web,clarke2010web,clarke2011web,clarke2012web}.  We use the
official binary subtopic judgments with \(\alpha=0.5\) and logarithmic discount
\(d_t=1/\log_2(t+1)\).  The 178 queries containing at least one positive document
assigned to multiple intents each contribute 50 pools of up to ten documents in
each of the three replicates.  Within each query, we cycle through its
overlapping documents as mandatory anchors and sample up to nine remaining
positive documents uniformly without replacement.  The anchor places every pool
outside the exact regime.  At \(K=5\), exact subset DP
provides the optimum for all \(178\times50\times3=26{,}700\) pools.  We compare
this optimum with marginal greedy, the prefix LP, and the projection bound.  We
then apply the LP at \(K=\min(20,n)\) to each complete positive-document pool.
For small pools, let \(L\) be the marginal-greedy value and \(U\) the
corresponding LP or projection upper bound.\footnote{For the projection bound,
\(L\) is the larger of the marginal-greedy value and the value of the ChainMerge
ranking of the projected instance, and \(U\) is the smaller of
\cref{eq:residual} and the same bound with every incidence deleted.}  Because subset DP supplies the
exact optimum, we report the certificate width as \((U-L)/\OPT\).  For complete
queries, where the exact optimum is unavailable, we report the certified
relative gap \((U-L)/U\).

The 20 complete queries with row support one also provide a real-data ERR-IA
test.  With binary relevance, the standard ERR mapping
\((2^{\mathrm{rel}}-1)/2^{\mathrm{rel}_{\max}}\) gives satisfaction \(0.5\) for
each positive judgment.  We use uniform intent priors, reciprocal-rank discount,
and \(K=\min(20,n)\), and compare ChainMerge with marginal greedy over 200
repetitions per query.

\paragraph{Leaderboard robustness.}
We collect the published mean \(\alpha\)-nDCG@20 of 124 official diversity runs
and form eight comparison groups matched by year, collection category, and
evaluated-topic count
\cite{clarke2009web,clarke2010web,clarke2011web,clarke2012web}.  Per-query run
scores are not published, so we bound them conservatively.  For query \(\ell\),
let \(L_\ell\) be the evaluator's greedy denominator and \(U_\ell\) the
prefix-LP upper bound.  The exact denominator lies in \([L_\ell,U_\ell]\), so
replacing \(L_\ell\) never raises a query score and lowers it by at most
\(\lambda_\ell=U_\ell/L_\ell-1\).  A run with published mean \(\sigma\) over
\(m\) topics therefore has an exact-denominator mean in
\([\sigma-\epsilon-\bar\lambda_m,\,\sigma+\epsilon]\), where \(\epsilon\) is
half a unit in the last published decimal and \(\bar\lambda_m\) averages the
\(m\) largest \(\lambda_\ell\) of that year.  An ordering
of two runs counts as preserved when their intervals are disjoint, so every
score consistent with the certificates keeps the published order.  All
calculations reproduce \texttt{ndeval}'s document-id tie-break.

\paragraph{Exact certificate arithmetic.}
We treat the evaluator's stored IEEE-754 binary64 logarithmic discounts as exact
dyadic rationals throughout subset DP, LP construction, scoring, and certificate
comparisons.  SoPlex~8.0.3 solves the LP in rational mode, and every returned
solution is rechecked for exact feasibility.  HiGHS supplies the numerical
solver times reported below.  For the projection bound, coordinate descent over
the retained-intent assignment \(\phi\) minimizes the deleted-incidence term
before the bound is evaluated exactly.

\paragraph{Runtime benchmarks.}
MIND-small assigns each news article one category and one subcategory
\cite{wu2020mind}.  Combining its train and development news tables gives
65,238 unique candidates, comprising 18 categories and 270 subcategories.  We
set item utility to \(1+\log(1+\text{training clicks})\), sort candidates once
by utility, and use prefixes of 5,000, 10,000, 20,000, 30,000, 45,000, and
65,238 candidates with \(K\in\{10,100,500\}\).  For graded \(\alpha\)-DCG with
\(\alpha=0.5\), we compare marginal greedy, which scans every remaining
candidate at each rank, with presorted ChainMerge, including its partition and
heap construction.  For binary gains, we additionally evaluate the count-only
path, which computes exact IDCG from group counts accumulated while judgments are
read.

For ERR-IA, we generate 65,238-candidate instances with either 18 or 270
intents.  Candidate-group frequencies and intent priors are drawn independently
from normalized lognormal weights, and satisfaction probabilities follow
Beta\((2,8)\).  Both methods receive the same instances and cutoffs.  Runtime
measurements use one CPU process and report the median of three repetitions.
Input parsing, utility sorting, and retrieval are shared preprocessing and are
excluded for both methods.

\FloatBarrier
\subsection{Results}

\paragraph{RQ1: exact recovery.}
ChainMerge attains the exhaustive optimum on all 3,000 generated
\(\alpha\)-DCG instances and all 3,000 generated ERR-IA instances.  On the 20
TREC queries with row support one, ChainMerge and marginal greedy also agree on
every ERR-IA objective, and ChainMerge is 7.2 times faster at the median.

\begin{figure}[t]
\centering
\includegraphics{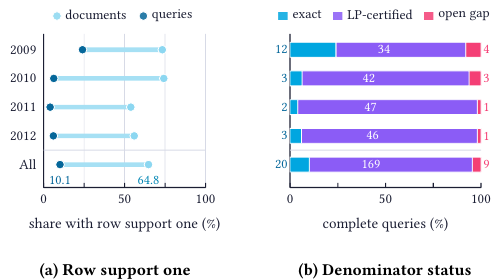}
\Description{Two panels sharing the rows 2009, 2010, 2011, 2012, and All. A
dumbbell chart compares the share of positive documents with row support one,
between 54 and 74 percent, with the share of complete queries, between 4 and 24
percent. A stacked bar chart splits each year's queries into exact,
LP-certified, and open-gap denominators.}
\caption{Official TREC Web 2009--2012 diversity judgments.  (a)~Share of
positive documents (light) and of complete queries (dark) with row support
one.  (b)~Status of
each complete query's \(\alpha\)-IDCG denominator at \(K=\min(20,n)\).  It is
exact by ChainMerge, certified optimal by the prefix LP, or left with a positive
certified gap.  Labels give query counts, and \cref{tab:trec,tab:yearcert} list
the exact values.}
\label{fig:trec}
\end{figure}

\paragraph{RQ2: prevalence.}
\Cref{fig:trec}a shows that 64.8\% of positive TREC documents have row support
one, but only 20 of 198 complete queries satisfy the condition throughout.
\emph{Document-level prevalence therefore does not carry over to queries.}  The
exact path applies directly to these 20 queries, and the remaining queries
require overlap certificates.  In MIND, category and subcategory are each
single-valued fields, so each induced judgment matrix has row support one
globally.

\paragraph{RQ3: overlap certificates.}
\emph{Small pools with exact optima.}
Across the 26,700 sampled overlapping pools, marginal greedy is strictly
suboptimal on 168 instances (0.63\%).  The prefix LP is a valid upper bound in
every instance and equals the exact subset-DP optimum on 26,692 (99.97\%).  In
the eight remaining instances, its gap never exceeds 0.56\%.  The projection
bound gives a zero-width certificate on 5,904 pools (22.1\%), with a mean width
of 7.4\%.

\emph{Complete-query certificates.}
On the 198 complete TREC queries, the LP certifies 189 greedy denominators as
globally optimal, including all 20 row-support-one queries and 169 overlapping
queries (\cref{fig:trec}b).  Across all queries, the mean certified gap is
0.0213\% and the maximum is 1.65\%.  The numerical LP solve takes 0.14 seconds
per query at the median.

\emph{Leaderboard robustness.}
The resulting run-level intervals preserve the winner in all eight matched
comparison groups and 1,069 of 1,074 strict pairwise orderings after accounting
for published rounding.  \emph{Replacing the greedy denominators with exact ones
therefore cannot change any group winner on these tracks.}

\paragraph{RQ4: runtime scaling.}
\begin{figure}[t]
\centering
\includegraphics{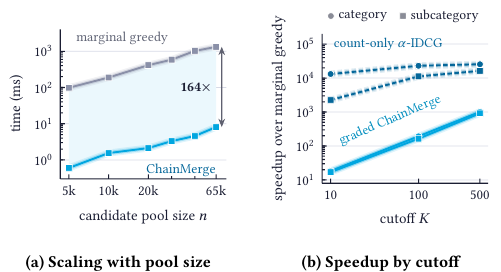}
\Description{Two log-log panels. Runtime of marginal greedy grows from about 100
to 1,300 milliseconds over pool sizes from 5,000 to 65,238, while ChainMerge
stays between 0.6 and 8 milliseconds. Speedup over marginal greedy grows with
the cutoff from about 17 to 1,011 for graded ChainMerge and lies between 2,250
and 25,427 for the count-only path.}
\caption{Runtime on MIND-small, as medians of three runs.  (a)~Graded
\(\alpha\)-DCG with the subcategory grouping (\(G=270\)) at \(K=100\) over six
candidate-pool sizes.  (b)~Speedup on the full pool of 65,238 candidates, for
graded ChainMerge over graded marginal greedy (solid) and for the count-only
\(\alpha\)-IDCG path over binary marginal greedy (dashed).  Circles mark the
category grouping (\(G=18\)) and squares the subcategory grouping (\(G=270\)),
as in (a).  \Cref{tab:runtime} lists the timings.}
\label{fig:runtime}
\end{figure}

Over the six pool sizes in \cref{fig:runtime}a, the runtimes show the
predicted separation between marginal greedy, whose cost grows with \(nK\), and
one partition followed by a small heap merge.  \Cref{fig:runtime}b shows how
the full-pool speedup changes with \(K\).  On the full MIND pool, graded ChainMerge is 17--1,011 times
faster across the two groupings and three cutoffs while matching every
marginal-greedy objective.  The count-only path is 2,250--25,427 times faster
than binary marginal greedy.  On the MIND-sized ERR-IA instances, ChainMerge
likewise matches every marginal-greedy objective and is 5--185 times faster
across the six intent-count and cutoff settings.  \emph{The speedup grows with
the cutoff because marginal greedy rescans the whole pool at every rank, while
ChainMerge pays for the pool only once.}

\section{Applicability and Limitations}

\paragraph{Use in evaluators.}
Row support can be checked as judgments are read.  When row support one holds,
ChainMerge returns the exact ideal denominator through the output-sensitive
path.  For an overlapping \(\alpha\)-DCG query, an evaluator can instead pair its
feasible ranking with the prefix-LP bound and report a certified upper bound on
its optimality gap.  \emph{Every reported \(\alpha\)-DCG denominator can thus
carry either an exactness guarantee or a certified gap.}  Across the eight matched TREC comparison groups, these
certified intervals preserve every winner and 99.5\% of strict run pairs after
published rounding.

\paragraph{Limits of the exact regime.}
The exact path requires row support one throughout a query, which held for only
20 of the 198 TREC queries.  Forcing a multi-intent document into a single
intent would change the objective rather than extend the exact regime, so the
remaining queries depend on certificates.

\paragraph{Limits of certification.}
For overlapping \(\alpha\)-DCG, the prefix LP upper-bounds the optimum and
recovers an optimal ranking only when an optimal \(x\) is integral.  Equality with a feasible value proves global
optimality, but when the certified gap is nonzero, the optimal ranking remains
unresolved.  The binary certificate extends to graded count-based decay through
the envelope of \cref{prop:graded}, still as an upper bound rather than an exact
algorithm.  Overlapping ERR-IA is not covered by this certificate.

\paragraph{Open complexity boundary.}
For \(\alpha\)-DCG, \cref{thm:alpha,thm:hard} establish tractability at row
support one and hardness at row support three, even when each intent occurs
twice.  Row support two remains open, as does the complexity parameterized by
the number of overlapping incidences or by the treewidth of the document--intent
incidence graph.

\section{Conclusion}

Evaluators have approximated ideal novelty rankings with marginal greedy
because computing them is NP-hard in general.  \emph{For \(\alpha\)-DCG, intent
overlap, rather than history-dependent gain itself, separates the tractable
cases from the hard ones.}  Under row support one, marginal greedy is exact for
both metrics, and ChainMerge attains the same optimum by merging virtual-score
chains instead of rescanning the pool.  Under overlap, the prefix LP bounds the
greedy \(\alpha\)-DCG denominator without constructing an ideal ranking,
certifying 189 of 198 official TREC denominators as optimal and leaving every
matched group winner unchanged.  Row support two and certification for
overlapping ERR-IA remain open.

\section{Ethical Considerations}

This work develops algorithms for ranking-metric computation and conducts no
user study or new data collection.  The experiments use publicly released TREC
judgments and MIND-small data.  \emph{Click counts are used only to construct runtime
workloads and support no behavioral or fairness conclusions.}

\appendix
\crefalias{section}{appendix}

\section{Detailed Results}\label{app:tables}

This appendix lists the values plotted in \cref{fig:trec,fig:runtime}.
\Cref{tab:trec} gives the row-support structure of the official TREC judgments,
\cref{tab:yearcert} the per-year prefix-LP certificates with their gaps and
solve times, and \cref{tab:runtime} the full-pool MIND-small timings behind
\cref{fig:runtime}b.  \emph{The certified gaps stay small in every year.  The
largest mean gap, 0.078\%, occurs in 2009, the year with the most open
queries.}

\begin{table}[H]
\caption{Structure of positive TREC Web diversity judgments.}
\label{tab:trec}
\centering
\footnotesize
\setlength{\tabcolsep}{3.2pt}
\begin{tabular}{lrrrr}
\toprule
& \multicolumn{2}{c}{Queries} & \multicolumn{2}{c}{Positive documents}\\
\cmidrule(lr){2-3}\cmidrule(lr){4-5}
Year & Total & Row support one & Total & Row support one\\
\midrule
2009 & 50 & 12 (24.0\%) & 4,942 & 3,622 (73.3\%)\\
2010 & 48 & 3 (6.3\%) & 6,553 & 4,862 (74.2\%)\\
2011 & 50 & 2 (4.0\%) & 5,030 & 2,707 (53.8\%)\\
2012 & 50 & 3 (6.0\%) & 5,559 & 3,111 (56.0\%)\\
\midrule
Total & 198 & 20 (10.1\%) & 22,084 & 14,302 (64.8\%)\\
\bottomrule
\end{tabular}
\end{table}

\begin{table}[H]
\caption{Complete-query prefix-LP certificates at \(K=\min(20,n)\). Gaps are
relative to the LP upper bound. Certificate decisions use exact rational
arithmetic, and times are numerical HiGHS medians.}
\label{tab:yearcert}
\centering
\small
\begin{tabular}{lrrrr}
\toprule
Year & Queries & Certified & Mean gap & Median sec.\\
\midrule
2009 & 50 & 46 & 0.0783\% & 0.084\\
2010 & 48 & 45 & 0.0001\% & 0.161\\
2011 & 50 & 49 & 0.0060\% & 0.132\\
2012 & 50 & 49 & 0.0002\% & 0.189\\
\midrule
Total & 198 & 189 & 0.0213\% & 0.137\\
\bottomrule
\end{tabular}
\end{table}

\begin{table}[H]
\caption{Full-pool MIND-small runtime (65,238 candidates), in milliseconds.
Panel~(a) includes partition and heap construction, and panel~(b) receives
counts accumulated during input.  Values are medians of three runs, and
speedups use unrounded times.}
\label{tab:runtime}
\centering
\small
\setlength{\tabcolsep}{2.8pt}
\begin{tabular}{lrrrrr}
\toprule
\multicolumn{6}{l}{\emph{(a) Graded \(\alpha\)-DCG, \(\alpha=0.5\)}}\\
Field & \(G\) & \(K\) & \shortstack{Marginal\\greedy} & ChainMerge & Speedup\\
\midrule
\multirow{3}{*}{Category}
 & \multirow{3}{*}{18} & 10  & 136.4 & 7.41 & 18.4\\
 & & 100 & 1,550.5 & 8.04 & 192.8\\
 & & 500 & 8,098.3 & 8.01 & 1,010.9\\
\midrule
\multirow{3}{*}{Subcategory}
 & \multirow{3}{*}{270} & 10  & 127.9 & 7.54 & 17.0\\
 & & 100 & 1,326.9 & 8.09 & 164.0\\
 & & 500 & 7,191.5 & 7.84 & 917.1\\
\midrule
\multicolumn{6}{l}{\emph{(b) Binary \(\alpha\)-IDCG from group counts, \(\alpha=0.5\)}}\\
Field & \(G\) & \(K\) & \shortstack{Marginal\\greedy} & Count-only & Speedup\\
\midrule
\multirow{3}{*}{Category}
 & \multirow{3}{*}{18} & 10  & 134.4 & 0.0102 & 13,120\\
 & & 100 & 1,520.0 & 0.0664 & 22,907\\
 & & 500 & 7,989.5 & 0.314 & 25,427\\
\midrule
\multirow{3}{*}{Subcategory}
 & \multirow{3}{*}{270} & 10  & 134.2 & 0.0597 & 2,250\\
 & & 100 & 1,339.4 & 0.120 & 11,161\\
 & & 500 & 6,682.8 & 0.411 & 16,250\\
\bottomrule
\end{tabular}
\end{table}

\FloatBarrier

\begin{acks}
The authors used AI tools to assist with manuscript preparation and code
development. All AI-generated content was reviewed and verified by the
authors, who take full responsibility for this work.
\end{acks}

\bibliographystyle{ACM-Reference-Format}
\bibliography{references}

\end{document}